\documentclass[runningheads]{llncs}
\usepackage[utf8]{inputenc} 
\usepackage[T1]{fontenc}    
\usepackage{hyperref}       
\usepackage{url}            
\usepackage{booktabs}       
\usepackage{amsfonts}       
\usepackage{nicefrac}       
\usepackage{microtype}      
\usepackage{lipsum}
\usepackage{newtxmath}
\usepackage{graphicx}
\usepackage{cite}
\usepackage{caption}
\usepackage{subcaption}
\usepackage{algorithm}
\usepackage{algorithmic}
\begin{document}
\title{Learning Large-scale Location Embedding from Human Mobility Trajectories with Graphs}
\titlerunning{GCN-L2V}
%
\author{Chenyu Tian\inst{1} \and
Yuchun Zhang\inst{2} \and
Zefeng Weng\inst{2} \and
Xiusen Gu\inst{2} \and
Wai Kin Victor Chan\inst{1}}
\authorrunning{C. Tian et al.}
%
\institute{Tsinghua-Berkeley Shenzhen Institute, Tsinghua Univeristy, Shenzhen 510006, China 
\email{tiancy19@mails.tsinghua.edu.cn}, \email{chanw@sz.tsinghua.edu.cn}\\
\and Tencent Inc, Shenzhen 510006, China\\
\email{\{yuchunzhang, mangozfweng,dennisgu\}@tencent.com}\\
}
\maketitle              
\begin{abstract}
An increasing amount of location-based service (LBS) data is being accumulated and helps to study urban dynamics and human mobility. GPS coordinates and other location indicators are normally low dimensional and only representing spatial proximity, thus diﬃcult to be effectively utilized by machine learning models in Geo-aware applications. Existing location embedding methods are mostly tailored for specific problems that are taken place within areas of interest. When it comes to the scale of a city or even a country, existing approaches always suffer from extensive computational cost and significant data sparsity.
Different from existing studies, we propose to learn representations through a GCN-aided skip-gram model named GCN-L2V by considering both spatial connection and human mobility. With a flow graph and a spatial graph, it embeds context information into vector representations. GCN-L2V is able to capture relationships among locations and provide a better notion of similarity in a spatial environment. Across quantitative experiments and case studies, we empirically demonstrate that representations learned by GCN-L2V are effective. As far as we know, this is the first study that provides a fine-grained location embedding at the city level using only LBS records. GCN-L2V is a general-purpose embedding model with high flexibility and can be applied in down-streaming Geo-aware applications.

\end{abstract}

\section{Introduction}
In modern society, location-based service (LBS) systems have brought great convenience to people's daily life with different specifications such as navigators, store recommendations, journey recording, pervasive games, and social networks. At the same time, LBS services generate a large amount of data every day. These data contain abundant offline behavior information. The mining of LBS data brings great potential for many industrial and commercial applications, such as traffic flow analysis, travel path recommendation, and location-based social networks.  
Location indicators like GPS coordinates are widely used in Geo-aware applications, but they are diﬃcult to be eﬀectively utilized by machine learning models because location indicators only contain the information of unique locations and physical distance. More than that, each location has rich information like land-use types, which are of vital importance to a wide range of Geo-related applications. In order to obtain meaningful and accurate information, a key step is to extract effective feature representations from the raw location indicator records.

Recently, unsupervised text encoding algorithms such as Word2Vec \cite{mikolov2013efficient} have been effectively utilized in many Natural Language Processing (NLP) tasks, and graph encoding algorithms such as node2vec\cite{node2vec} have been widely applied in graph learning tasks. Word embedding and graph embedding greatly boosted related studies. The intuition behind that is training deep models which encode words or nodes into vector space representations based on their positions and contexts. Similar ideas can be borrowed into the field of Geographic Information Science (GIS). 
Finding the latent representations of different locations means deriving an embedding from high-dimension sparse location patterns (like the one-hot encoding) to low-dimension dense patterns, which can serve as the geographical "Word2Vec" to help Geo-related studies. If there are models that can map each location into a meaningful embedding vector, they can benefit nearly all studies that take spatial information as inputs, such as similar position querying and place recommendations.

However, there are few available large-scale location embedding products or algorithms. Existing location embedding methods mostly tailored for speciﬁc problems that are taken place within Areas of Interest (AOI) and Point of Interest (POI) \cite{yao2017serm, feng2018deepmove, liu2016exploring, yan2017itdl}. In reality, the quantity of locations without specific utility (like suburban areas without POIs) is much larger, but these locations are ignored in most studies. These limitations constrain the utility and flexibility of these models. Also, when it comes to the scale of large cities, countries, or even the whole earth, existing approaches may suﬀer from extensive computational cost and signiﬁcant data sparsity.

The motivation of this study is to learn meaningful location embedding using LBS data. Because these data contains offline human behaviour, which reveals important association information among locations through the dynamic human mobility flow. In addition, human mobility data generally have better coverage of unpopular areas and time periods. It means analyzing human mobility has the potential to mine urban patterns through location embedding. To find latent representations of Geo indicators, the key challenge lies in how to generate a meaningful context for locations using the mobility flow data, similar to the sentence contexts for word embedding or the neighbor contexts for graph embedding. Another challenge is the data sparsity because the records usually follow a long-tail distribution with regard to locations\cite{yan2017itdl}.
This study leverage the massive amount of raw LBS records to train a location embedding. We address this issue by proposing a method that generates ﬁne grained location embedding, which leverages spatial information and human mobility trajectories.
We implement quantitative experiments and case studies, which empirically demonstrate that the representations learned by the proposed model are effective.
In summary, the key contributions of this work are:
\begin{enumerate}
    \item We study a graph-aided representation learning model called GCN-L2V (GCN aided Location2Vec) for learning general-purpose location embedding. To leverage both human mobility and spatial information, we integrate flow graph and spatial graph, and incorporate both contexts in an unsupervised manner. 
    \item To the best of our knowledge, this is the first location representation learning model that combines both GCN and word embedding techniques, which complement each other and derive the vector representations under large-scale circumstances. Also, GCN-L2V has high flexibility, and new relations(or graphs) can be imported easily.
    \item As far as we know, this is the first study that provides a fine-grained location embedding for every part in large cities using only LBS records, and we design quantitative experiments and case studies to illustrate a better notion of semantic similarity in a spatial environment and empirically demonstrate the effectiveness.
\end{enumerate}

\section{Related Works}
Learning the embedding of places and locations has been a popular research topic in the urban computing field. In recent years, several efforts have been made to encode GPS coordinates or POIs at the feature level.
In many supervised learning studies, location embeddings are indirectly trained as an auxiliary module to import spatial information to perform classification and regression tasks. Feng \textit{et al.} \cite{feng2018deepmove} proposed DeepMove to predict human mobility using a multi-modal embedding recurrent
neural network and location embedding is correspondingly generated as a byproduct of the original prediction-based model. Gao \textit{et al.} \cite{gao2017identifying} identified and linked trajectories to users. The model represents each check-in with a low-dimensional vector to mitigate the problem of the curse of dimensionality. The intuition of location embedding is also applied in various tasks including next location prediction \cite{yao2017serm}, place-of-interest recommendation \cite{chang2018content}, and trajectory similarity computation \cite{li2018deep}. However, these methods are mostly tailored for speciﬁc problems that are taken place within areas of interest, while information of less-visited locations has hardly been cared about. When it comes to the scale of large cities, countries, or even the whole earth, existing approaches always suﬀer from extensive computational cost and signiﬁcant data sparsity. 

Some studies directly learn location embeddings. Yin \textit{et al.} \cite{yin2019gps2vec} trained a neural network in each Universal Transverse Mercator (UTM) zone to learn the semantic embeddings for Geo-coordinates worldwide. The training labels are derived from large-scale geotagged documents such as tweets, check-ins, and images that are available from social sharing platforms. Mai \textit{et al.} \cite{mai2020multi} proposed Space2vec using an encoder-decoder framework to encode the absolute positions and spatial relationships of places based on POI information. 
But for large-scale location embedding studies, labels for all locations are expensive to get and the attributes are difficult to define because some locations may have multiple labels and unpopular locations have no label. Also, the labels of the locations may change over time.

Compared with labeled location data like POI types, mobility trajectories are more easily accessible. Recent studies have applied the embedding methods to mobility data with more spatio-temporal details. Models have applied ideas similar to word embeddings in NLP using social media check-in data, where an analogy of locations = words, trajectories = sentences was made \cite{poi2vec, liu2016exploring, yan2017itdl,zhou2018deepmove}. The drawback of these models is that the local context window approach and negative sample sampling ignore the overall relationship and cause data sparsity for less-visited locations. Shimizu \textit{et al.} \cite{shimizu2020learning} leveraged spatial hierarchical information according to the local density of observed data points, and generated fine-grained place embeddings using human mobility trajectories. ZE-Mob created embeddings of places using the New York Taxi GPS dataset \cite{yao2018representing}. Wang and Li \cite{wang2017region} considered both temporal dynamics and multi-hop transitions in learning the dynamic region representations and proposed to jointly learn the representations from a flow graph and a spatial graph. However, they are mostly tailored for specific problems that are taken place within a small amount of AOI and POI, while ignoring the majority of unpopular locations.



\section{Method}

\subsection{Problem Formulation}
In this study, the training material consists of a huge amount of LBS records, each record has the format as $(user ID, time, latitude, longitude)$. Given the human mobility records, we aim to learn a vector representation $u_l \in \mathbb{R}^d$ in $d$-dimensional embedding space.
\subsubsection{Space Discretization}
\begin{figure}
  \centering
  \includegraphics[width=.5\textwidth]{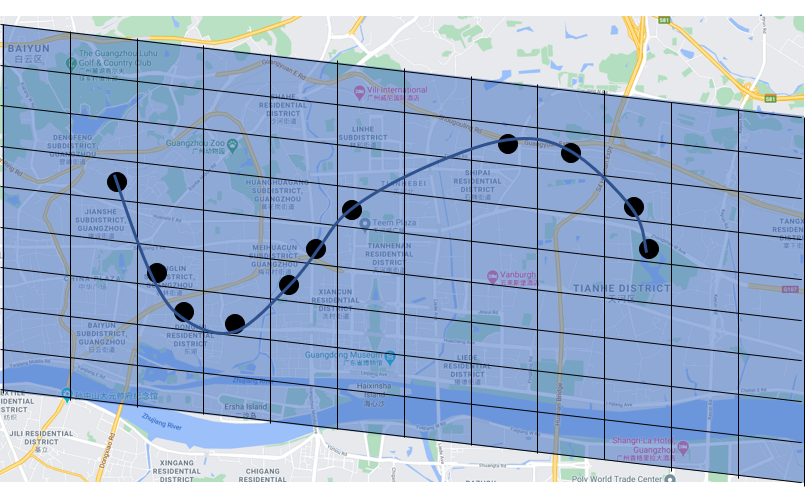}
  \caption{Split the area into locations using Google S2, then project the LBS data records into location trajectories.}
  \label{fig:fig1}
\end{figure}

In practice, it is difficult to embed infinite continuous location indicators.
A more practical way is to split continuous Geo indicators into discrete cells. Instead of splitting locations into grids as \cite{shimizu2020learning, yin2019gps2vec}, our study uses Google S2 geocoding, which is a public domain geocoding algorithm that uses a Hilbert space-filling curve to split the earth surface into hierarchical cells. Google S2 geocoding offers properties like arbitrary precision, fractal curve continuity, and homogeneous cell size. In this way, we encode the continuous geographical coordinates into detailed discretized cell representations, referred to as "locations" in this paper (as the cells shown in the Figure \ref{fig:fig1}).
Human mobility trajectories are thus extracted as sequences of locations as follows. For each user, stay points are firstly converted into corresponding location IDs by space discretization. Trajectories are then generated by partitioning the records with a maximum time window between consecutive records to ensure good correlations. The extracted trajectories serve as the context information in the representation learning task below.

\subsubsection{Location Embedding}
We denote the embedding of location $l$ as $u_l$, which is a $d$-dimensional vector. Similar to word embedding, location embedding is a vector representation of the characteristics of locations learned from human mobility trajectories. The location embedding should be able to reflect the similarity among locations with highly-related locations close to each other in the vector space.

\subsection{Graph Construction}
There are two kinds of relations we want to capture, mobility flow relationship and spatial adjacency. The first type of relationship is derived from the mobility flow among the locations, which is formulated as a flow graph $G_f$. The second one is the spatial adjacency defined as a spatial graph $G_s$. The intuitions and definitions of these two graphs are introduced in detail as follows.

\subsubsection{Flow Graph}
If two locations frequently co-occur in human trajectories, they are more likely to have correlations. For example, people usually travel among different subway stations. So we define the flow graph as a graph $G_f=(V, E_f)$, where vertices $V$ is the location set and $E_f$ represents edges. The edge weight $e_{ij}$ on $E_f$ represents the frequency of co-occurrence of location $i$ and location $j$ in consecutive manner.

\subsubsection{Spatial Graph}
Each path on the flow graph consists of a sequence of locations, and the flow graph models the mobility pattern of crowds. However, there are some issues with the flow graph. During learning the representation of locations, the flow graph suffers from data sparsity, which means that if there is no mobility flow going in or out of certain locations during the studied time span, the information of this region may be deficient. Secondly, the flow graph cannot recognize nearby locations. Thus, a spatial graph is introduced.

The basic assumption of the spatial graph is that human mobility is bounded by space. Typically, adjacent locations are more likely to have similar characteristics (like locations in the same natural parks and residential neighborhoods). When there is no location transition observed, the probability that people appeared at a different location is inversely correlated to the distance they need to travel.
The spatial graph $G_s=(V, E_s)$ shares the same structure and exactly the same vertices as the flow graph. The only difference is the edge construction mechanism. The edge weight $e_{ij}$ on $E_s$ refers to the adjacency between location $i$ and location $j$, which is defined as 
\begin{equation}
    e_{ij} = \exp \left(-\frac{\operatorname{dist}\left(i, j\right)}{\Delta}\right), \text { if dist }\left(i, j\right) \leq \Delta,
\end{equation}where $\operatorname{dist}(i, j)$ is the spatial distance between the centers of the two locations, and $\Delta$ is the maximum distance threshold.
This design of the spatial graph naturally incorporates spatial adjacency.

\subsection{GCN aided Embedding Layer}
Graph Convolution Network (GCN) has achieved extraordinary performance on several different types of tasks with the graph structure, such as node classification and network representation. GCN manipulates the spectral domain with graph Fourier transforms to apply convolutions in spectral domains \cite{bruna2013}. The following study makes graph convolution more promising by reducing the computational complexity from $\mathcal{O}(n^2)$ to linear \cite{kipf2016semi}.

\begin{figure}
  \centering
  \includegraphics[width=.8\textwidth]{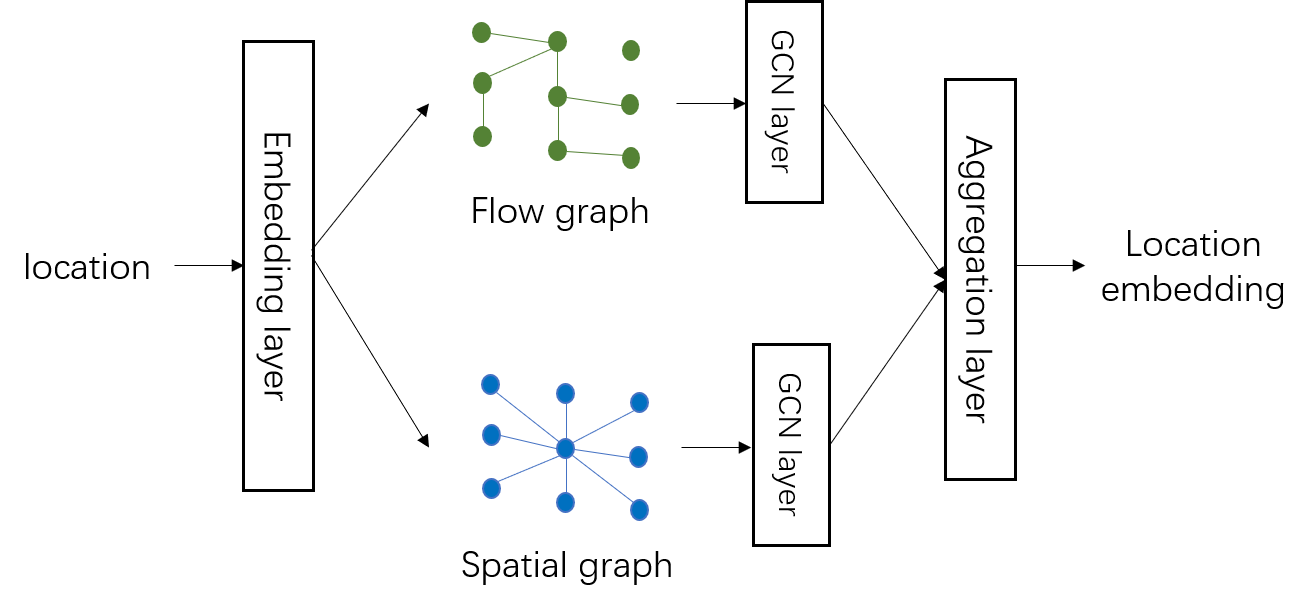}
  \caption{The framework of the GCN-L2V.}
  \label{fig:geoembedding}
\end{figure}
In order to include these two graphs mentioned above, our proposed GCN-L2V model learns the representation of each location shown as Figure \ref{fig:geoembedding}.
First, there is an embedding layer with the embedding matrix $U^0$, which maps every location $l$ to a vector representation $u^0_l$. After that, parallel GCNs are applied to aggregate information from neighbor nodes on the flow graph and the spatial graph. Without loss of generality, mathematically the computation of GCN on flow graph follows this formula:
\begin{equation}
    U^f=\sigma\left(\tilde{D}_f^{-\frac{1}{2}} \tilde{A}_f \tilde{D}_f^{-\frac{1}{2}} U^0 W_f^{(l)}\right),
    \label{gcn}
\end{equation}
where $\sigma$ is the non-linear activation function, and $W_f$ is the weight matrix of the GCN layer. $\tilde{D}_f$ and $\tilde{A}_f$ represent normalized degree matrix and adjacency matrix of the flow graph.
Following, an aggregate layer is applied to derive the final embedding as
\begin{equation}
    U=Agg(U^f, U^s),
\end{equation}
where $U^f$ and $U^s$ are the embedding learned by flow graph and spatial graph computed as Equation \ref{gcn}, $Agg(\cdot)$ can be aggregation operations like mean/max pooling.
Instead of flow graph and spatial graph, it should be noticed that our proposed model has high flexibility and other types of graph can be imported in parallel.

\subsection{Representation Learning}
\begin{figure}
  \centering
  \includegraphics[width=.9\textwidth]{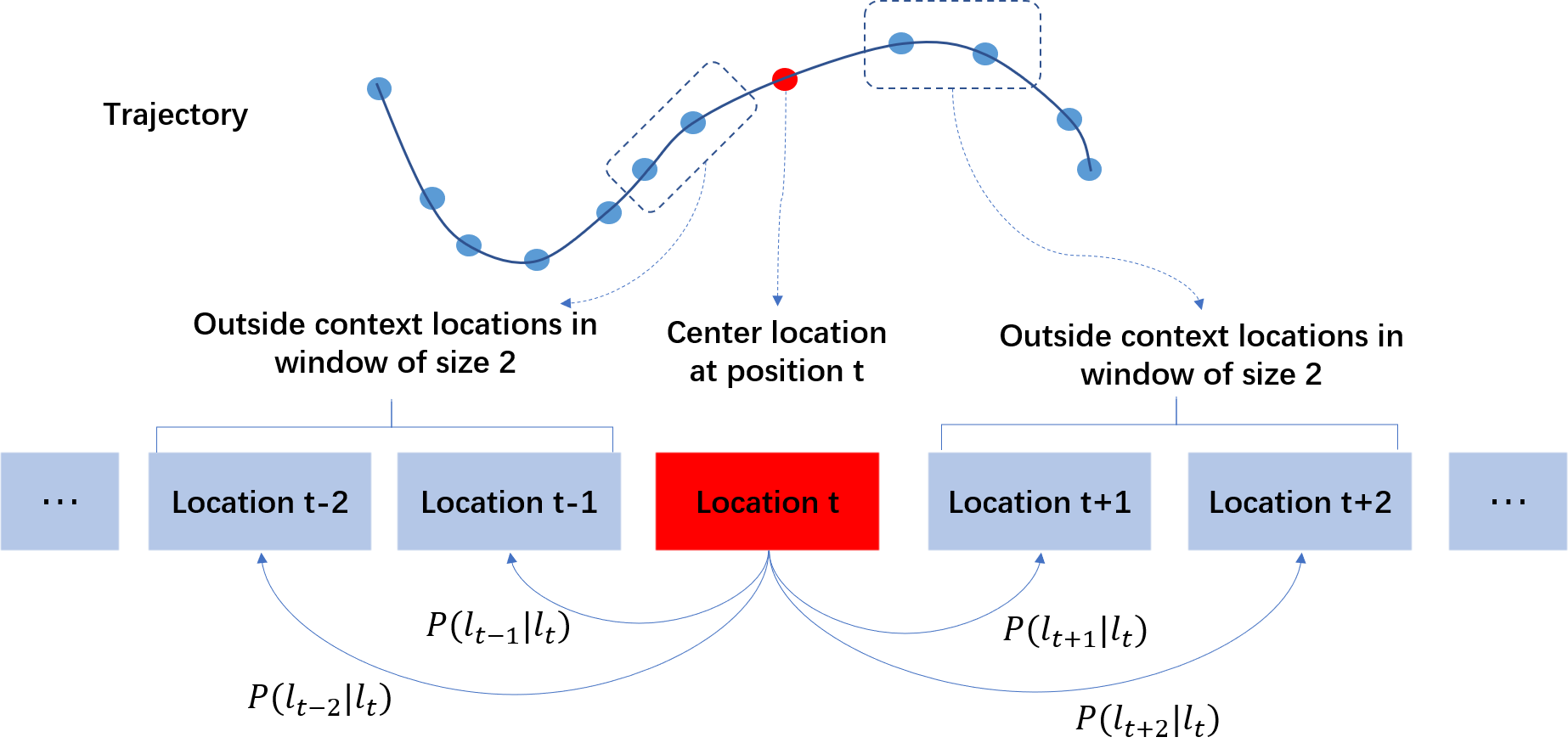}
  \caption{Skip-gram model for learning location embedding.}
  \label{fig:learning}
\end{figure}
We adopt the skip-gram model \cite{mikolov2013distributed} to learn location embedding. As shown in Figure \ref{fig:learning}, for every trajectory $\tau$ as $(l_1, l_2, ... , l_n)$, we go through each position in the trajectory and define the center location as $l_c$ and its context locations as $l_o$. To identify the context locations of the $t$-th position, define a window of size $m$ which means the model will take locations from $t-m$ to $t+m$ as the context. We try to maximize the likelihood of the context locations given the center location. This likelihood can be represented using the formula,
\begin{equation}
    L(\theta)=\prod_{t=1}^{n} \prod_{-m \leq j \leq m \atop j \neq 0} P\left(l_{t+j} \mid l_{t} ; \theta\right),
\end{equation}
where $\theta$ is the parameters of the model, and
\begin{equation}
    P(l_o \mid l_c; \theta)=\frac{\exp \left(\boldsymbol{u}_{l_o}^{'\top} \boldsymbol{u}_{l_c}\right)}{\sum_{l \in \text { Locations }} \exp \left(\boldsymbol{u}_{l}^{'\top} \boldsymbol{u}_{l_c}\right)},
\end{equation}
where ${u}'_{l}$ and ${u}_{l}$ are the context and node vector representations of location $l$.
To put this equation in a form that it is easy to take derivatives, take the log of the equation. Also, negative sampling \cite{mikolov2013distributed} is used with $K$ as the number of negative samples.
The likelihood function can be formulated as
\begin{equation}
    L\left(\theta \right)=\sum_{t=1}^{n}\left(\sum_{-m \leq j \leq m \atop j \neq 0} \log \left(\sigma\left({\boldsymbol{u}_{l_{t}}^{\top}\boldsymbol{u}'_{l_{t+j}}} \right)\right)-\sum_{k=1}^{K} \log \left(\sigma\left(\boldsymbol{u}_{l_{t}}^{\top} \boldsymbol{u}'_{z_k}\right)\right)\right), 
    \label{skipgram}
\end{equation}
and $z_k$ are sampled locations. Finally, learn the embedding by maximizing the probability of real context locations and minimizing the probability of random sampled locations. Overall, the whole procedure is summarized as Algorithm \ref{algorithm}.
\begin{algorithm}
	\renewcommand{\algorithmicrequire}{\textbf{Input:}}
	\renewcommand{\algorithmicensure}{\textbf{Output:}}
	\caption{GCN-L2V}
	\begin{algorithmic}[1]
    	\REQUIRE LBS records, number of epochs $T$;
		\STATE Preprocess LBS records into location trajectories $(\tau_1, \tau_2, ..., \tau_N)$;
		\STATE Construct graph $G_f$, $G_s$
		\STATE Initialize parameters $U^0, U^{0'}, W_f, W_s$;
		\WHILE{$t < T$} 
		\FOR{trajectory $\tau$ in $(\tau_1, \tau_2, ..., \tau_N)$}
		\FOR{location in $\tau$}
		\STATE Record context locations;
		\STATE Sample negative locations;
		\ENDFOR
		\STATE Compute likelihood function according to Eq.\ref{skipgram};
		\STATE Update parameters to maximize likelihood by gradient ascend;
		\ENDFOR
		\IF{converge}
		\STATE break;
		\ENDIF
		\ENDWHILE 
		\ENSURE location embedding $U$.
	\end{algorithmic}
	\label{algorithm}
\end{algorithm}

\section{Experiment}
The following section describes the dataset and the strategies used to evaluate the proposed method. We perform two quantitative evaluations against the embedding baselines. Afterwards, a down-stream task and case studies are carried out to illustrate the effectiveness of the embedding learned.
\subsection{Dataset Description}
We examine the performance of our proposed model on a real-world dataset of Guangzhou, which is one of the most populous cities in China with an area of 7434 $km^2$. The data are collected from 1st December to 31st December 2019. The overall number of LBS records is 75,132,685 collected from 1,763,585 users. After prepossessing, we map the GPS coordinates into 109,455 locations, with Google S2 level in 16 (the average area of each location is about 19,800 $m^2$). 4,322,756 edges are constructed in the spatial graph with a maximum distance threshold equals to 500$m$ and 864,752 in the flow graph.
\subsection{Model Evaluation}
A well-designed location embedding should: (1) preserves semantic and spatial information at the same time; (2) encodes similar (dissimilar) locations to similar (dissimilar) embedding vectors, where the similarity between two vectors can be characterized by some metrics, such as cosine similarity score; (3) can serve as inputs for down-streaming tasks and improve the performance. To demonstrate the effectiveness of the proposed model, we compare it with the following representation learning baselines:
\begin{enumerate}
    \item \textbf{Word2Vec}\cite{mikolov2013efficient} is an architecture that can be used to learn embeddings from large datasets. Embeddings learned through Word2Vec have proven to be successful on a variety of downstream natural language processing tasks. It is adapted to location representation learning task by treating locations as words and trajectories as sentences.
    \item \textbf{Node2vec}\cite{node2vec} is an algorithmic framework for representation learning on graphs. Given any graph, it can learn continuous feature representations for nodes, which can then be used for various downstream machine learning tasks.
    \item \textbf{LINE}\cite{tang2015line} is a graph embedding method that learns embedding on a weighted graph to encode both first and second order proximity with LINE(1st) represented for the former and LINE(2nd) represented for the latter.
    \item \textbf{GVAE}\cite{gvae} is a framework for unsupervised learning on graph-structured data based on the variational auto-encoder by making use of latent variables and is capable of learning interpretable latent representations for undirected graphs.
\end{enumerate}
In our experiments, Node2vec, LINE, and GVAE are trained on a graph which consists of both the flow graph and the spatial graph. As ablation experiments, GCN-L2V (flow w/o) is the GCN-L2V without using the flow graph, and GCN-L2V (spatial w/o) is without using the spatial graph.

To test the effectiveness of our proposed algorithm, we design the following tasks to evaluate the ability of semantic and spatial information representation separately, and apply the location embedding in a down-streaming task.

\subsubsection{Urban Subway Station Detection}
\begin{figure}
  \centering
  \includegraphics[width=.9\textwidth]{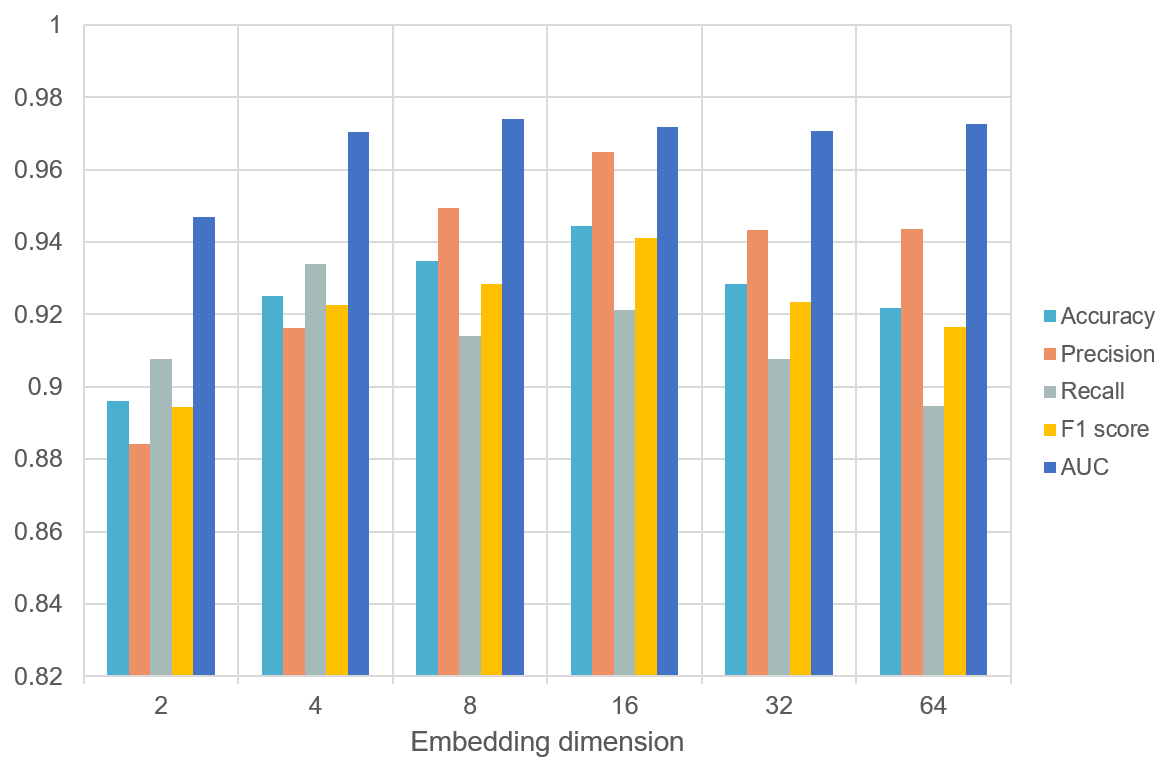}
  \caption{The metrics of GCN-L2V on different dimensions.}
  \label{cmps}
\end{figure}

As mentioned, location embedding should contain semantic information of locations. Thus, we firstly use the learned location embedding directly as features in a classification model to detect subway stations. In practice, we use the logistic regression as the classifier with 5-fold cross-validation. In terms of the experimental settings, we sample an equal number of subway station locations and non-subway locations. Then the goal is to learn a binary classification model to determine whether a given location is a subway station.
The measurement metrics are accuracy, precision, recall, F1 score, and Area Under Curve (AUC). First, we try different embedding dimensions. With the increase of the embedding dimension, the performance of GCN-L2V increases gradually as shown in Figure \ref{cmps}. Slight over-fitting occurs after the dimension is over 16. Thus, the embedding dimension is set as 16 in the following.

As shown in Table.\ref{t1}, all embeddings learned by these methods are more effective compared with raw GPS coordinates (the metrics for GPS are correspondingly 0.5483, 0.5350, 0.5841, 0.5545, and 0.5632). LINE(1st) performs worst in these embedding methods. LINE(2nd) is slightly better than Node2Vec, Word2Vec, and GVAE. The embedding learned by GCN-L2V results with best performance. Moreover, GCN-L2V (spatial w/o) also performs better than GCN-L2V (flow w/o), which means flow graph is more important in learning context information compared with spatial graph. Complementing with each other, GCN-L2V achieves satisfactory results and learns the patterns of subway stations well.

\begin{table}[]
\caption{The metrics of different methods on binary classification}
\centering
\begin{tabular}{l|lllll}
\hline
Models               & Accuracy        & Precision       & Recall          & F1 score        & AUC             \\ \hline
Word2Vec             & 0.8959          & 0.9215          & 0.8626          & 0.8903          & 0.9238          \\
Node2Vec             & 0.8896          & 0.8915          & 0.8884          & 0.8865          & 0.9442          \\
LINE(1st)           & 0.8538 & 0.9024   & 0.7978  & 0.8382   & 0.9086\\
LINE(2nd)  & 0.9122     &0.9534   & 0.8692  & 0.9055 &  0.9474\\
GVAE  & 0.8893    &0.9149   & 0.8559  & 0.8792 &  0.9308\\
GCN-L2V              & \textbf{0.9445} & \textbf{0.9650} & \textbf{0.9211} & \textbf{0.9410} & \textbf{0.9718} \\ \hline
GCN-L2V (flow w/o) & 0.8895          & 0.8939          & 0.8819          & 0.8860          & 0.9472           \\
GCN-L2V (spatial w/o)    &         0.8960   &   0.9158  &         0.8757   &    0.8917     &     0.9537    \\ \hline
\end{tabular}
\label{t1}
\end{table}

\subsubsection{Region Analysis}
Except for semantic information, spatial information should also be kept. Here, we define "regions" as larger areas with specific land-use types, for example, schools, hospitals, transport hubs, and residential communities. Some regions are of comparatively large areas, thus made up of multiple locations. Intuitively, the embeddings of locations in the same region should be similar to each other. In this task, we use region boundary data with irregular shapes. It contains more than 10,000 regions in Guangzhou. We first sample one location in each of these regions. Then, we find its closest $K$ locations accordingly in terms of cosine similarity as Equation \ref{cossim}. 
\begin{equation}
    \operatorname{Similarity}<\boldsymbol{u}_i, \boldsymbol{u}_j>=\frac{\boldsymbol{u}_i^{\top} \boldsymbol{u}_j}{\|\boldsymbol{u}_i\| \cdot \|\boldsymbol{u}_j\|}
    \label{cossim}
\end{equation}
We hope that these similar locations still in the same region. Therefore, we use the top $K$ accuracy (Accuracy@K) as the evaluation metric, which calculates the percentage of top $K$ closest locations that are still in the same region. As shown in Table \ref{t2},  LINE(2nd) has the worst performance on region analysis although it is the second best model in the classification task. The embeddings learned by GCN-L2V still result with best performance. From the comparisons, it can be found that adding spatial graph has more significance in improving the performance than adding flow graph, because spatial graph contains overall distance information between locations.

\begin{table}[]
\caption{The metrics of different methods on region analysis}
\centering
\begin{tabular}{l|lll}
\hline
Models               & Accuracy@5 & Accuracy@10 & Accuracy@20     \\ \hline
Word2Vec             & 0.5615     & 0.4523      & 0.3635         \\
Node2Vec             & 0.6077   & 0.5254 &  0.4373         \\
LINE(1st)            & 0.5385 & 0.4151 &  0.3324         \\
LINE(2nd)         & 0.4692   & 0.3617 &  0.3318     \\
GVAE         & 0.8631  & 0.7823 & 0.6623 \\
GCN-L2V              & \textbf{0.8754}  & \textbf{0.8586}   & \textbf{0.6681} \\ \hline
GCN-L2V (flow w/o)   & 0.8646    &0.8015    &     0.6604            \\
GCN-L2V (spatial w/o) &  0.7077  &  0.7314   &    0.4893   \\ \hline
\end{tabular}
\label{t2}
\end{table}

\subsubsection{User Risk Behavior Classification based on Trajectories}
Based on the former two experiments, the embedding learned by GCN-L2V achieves a good balance between semantic and spatial characterization. For real-world applications, location embedding can serve as input features for down-streaming tasks. In our industry practices, we collected risk behavior labels and LBS records from 46,873 anonymous users. Their online risk behaviors are labeled with "0" for normal and "1" for high-risk. About 12.18\% of them are labeled with high risk. There are over 3 billion LBS records, and we sample the first 100 records for each user (padding is applied for users with fewer than 100 records). This down-streaming task aims to detect high-risk groups solely rely on their trajectories.
For the experimental setup, we use a LSTM network as the classifier, and the input data are location embedding together with time-related attributes, i.e. hour and day of the week.
\begin{table}[]
\caption{The metrics of different methods on credit risk classification}
\centering
\begin{tabular}{l|l}
\hline
Models               & AUC     \\ \hline
raw GPS & 0.5460\\
Word2Vec             & 0.5748      \\
Node2Vec             & 0.5851 \\
LINE(1st)            & 0.5809 \\
LINE(2nd)          & 0.5816 \\
GVAE         & 0.5877 \\
GCN-L2V              & \textbf{0.5933}\\ \hline
GCN-L2V (flow w/o)   & 0.5830 \\
GCN-L2V (spatial w/o) &  0.5892  \\ \hline
\end{tabular}
\label{t3}
\end{table}
Based on Table \ref{t3}, location embedding learned by GCN-L2V can improve the classification accuracy by about 8\% compared with GPS sequence as inputs. The improvement is satisfactory considering the difficulty of this task only using LBS records while all other online attributes are absent. This experiment tells that mining the offline mobility behavior helps to depict human profiles and discover useful features for applications.

Overall, GCN-L2V achieves the best performance on all these tasks. 

\subsection{Case Study}
To better illustrate the location embedding's ability in capturing locations' relationships via human mobility, some representative locations (labeled as red stars) and their most related locations (labeled as black dots) in terms of cosine similarity in the vector space are presented.

\begin{figure}
     \centering
     \begin{subfigure}[t]{0.3\textwidth}
         \centering
         \includegraphics[width=\textwidth]{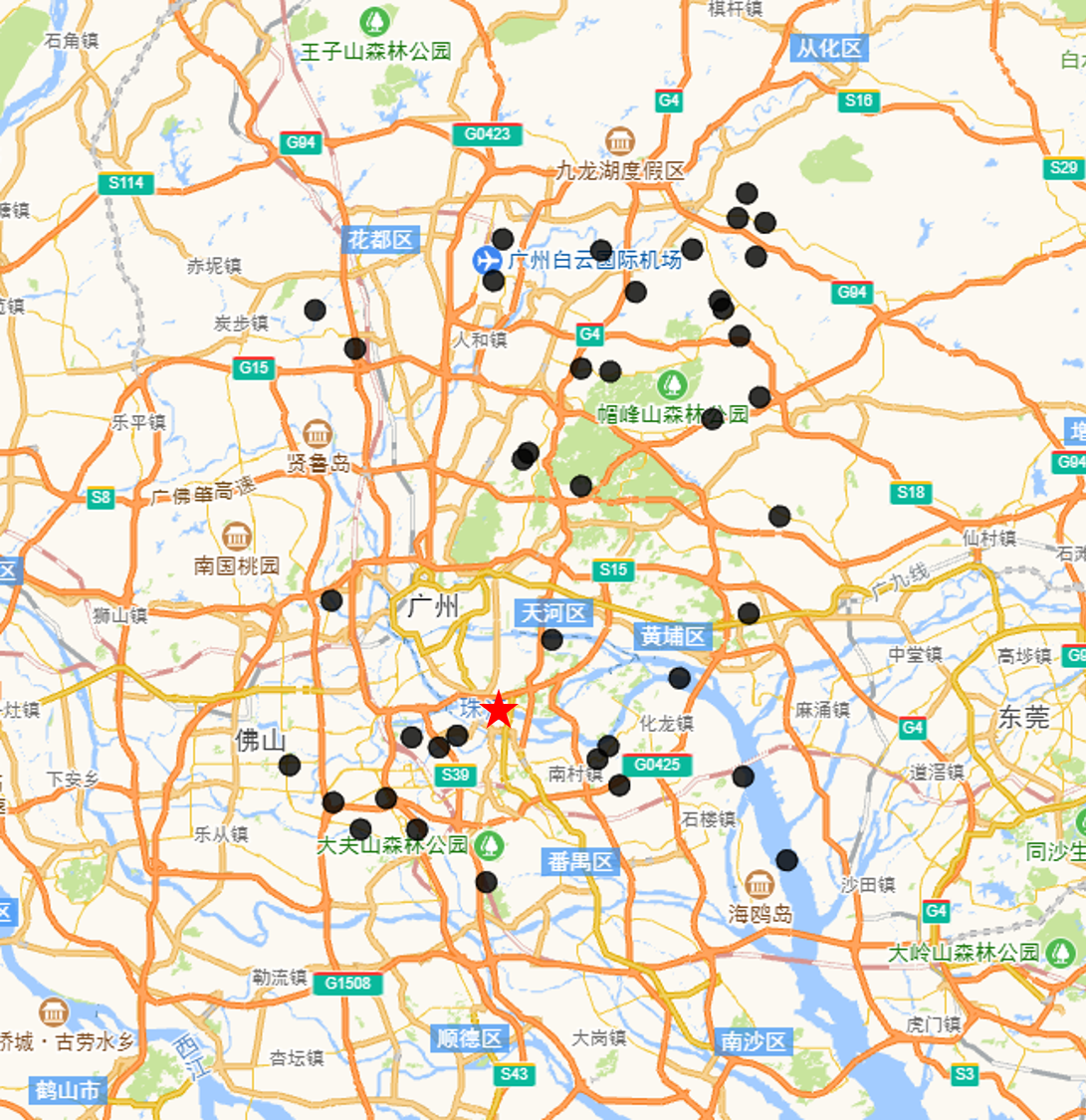}
         \caption{Word2Vec.}
     \end{subfigure}
     \hfill
     \begin{subfigure}[t]{0.3\textwidth}
         \centering
         \includegraphics[width=\textwidth]{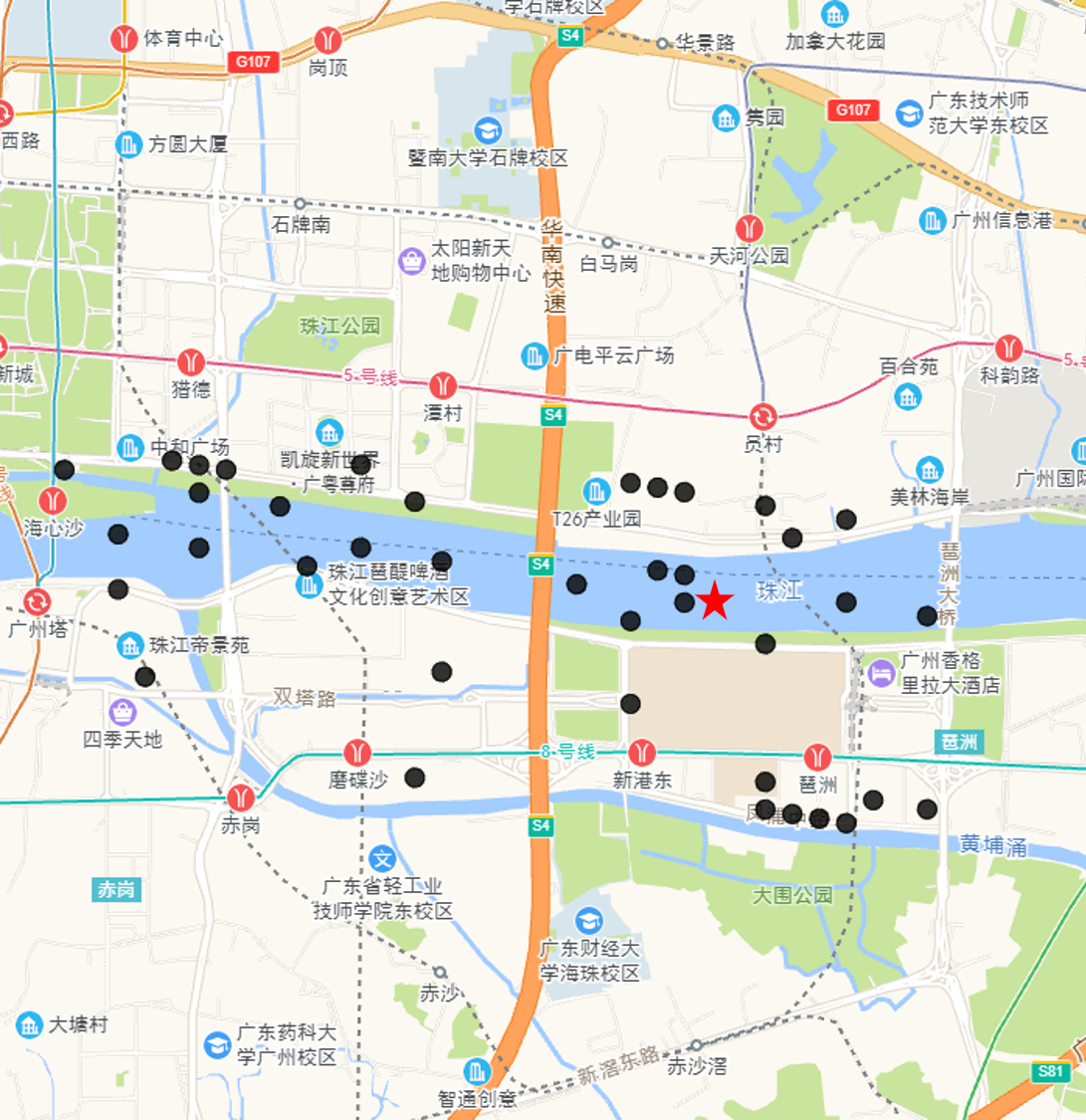}
         \caption{Node2Vec.}
     \end{subfigure}
     \hfill
     \begin{subfigure}[t]{0.3\textwidth}
         \centering
         \includegraphics[width=\textwidth]{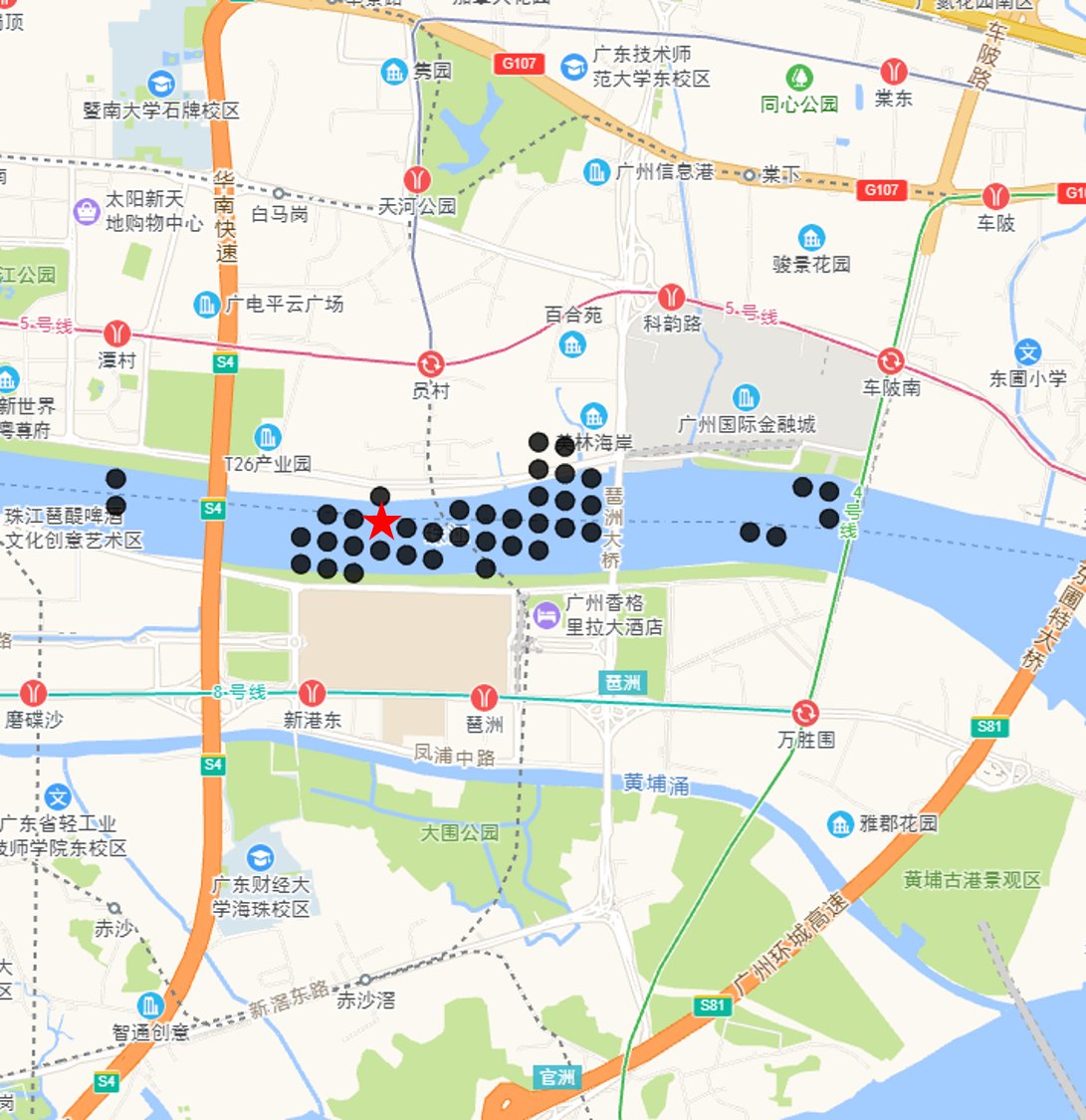}
         \caption{GCN-L2V.}
     \end{subfigure}
        \caption{Unpopular locations and their neighbours in latent space.}
        \label{unpopular}
\end{figure}
As mentioned before, a large number of unpopular locations should not be ignored. As shown in Figure \ref{unpopular}, select a location in the river as the center location, which is rarely visited because of geographical constraints. For Word2Vec, the learned related locations distribute randomly in the whole city. The performance is unacceptable because there are few contexts for training this location and the information loss is severe. For Node2Vec, because every node gets an almost equal number of contexts for training by random walking on the graph, the related locations are close to the center location. However, the embedding cannot distinguish the river boundary with some related locations on the land and others on the water. In comparison, the embedding learned by GCN-L2V is more reasonable, and most of the related locations distribute along the river without exceeding the river boundary.
Apart from negative sampling, GCN-L2V solves the data sparsity issue by importing graphs. In the embedding layer, these graphs bring background information of human mobility and spatial adjacency. Also, both the context and node vector representations are processed by GCNs. Because every location has connected edges on these graphs, the learning of location embedding always considers their neighbors. Thus, all locations get exposed during training directly as subjects or indirectly as neighbors. In this way, GCN-L2V not only contains the dynamic properties of the skip-gram model but also tackles the data sparsity issue by importing global relations via GCNs.

\begin{figure}
     \centering
     \begin{subfigure}[t]{0.45\textwidth}
         \centering
         \includegraphics[width=\textwidth]{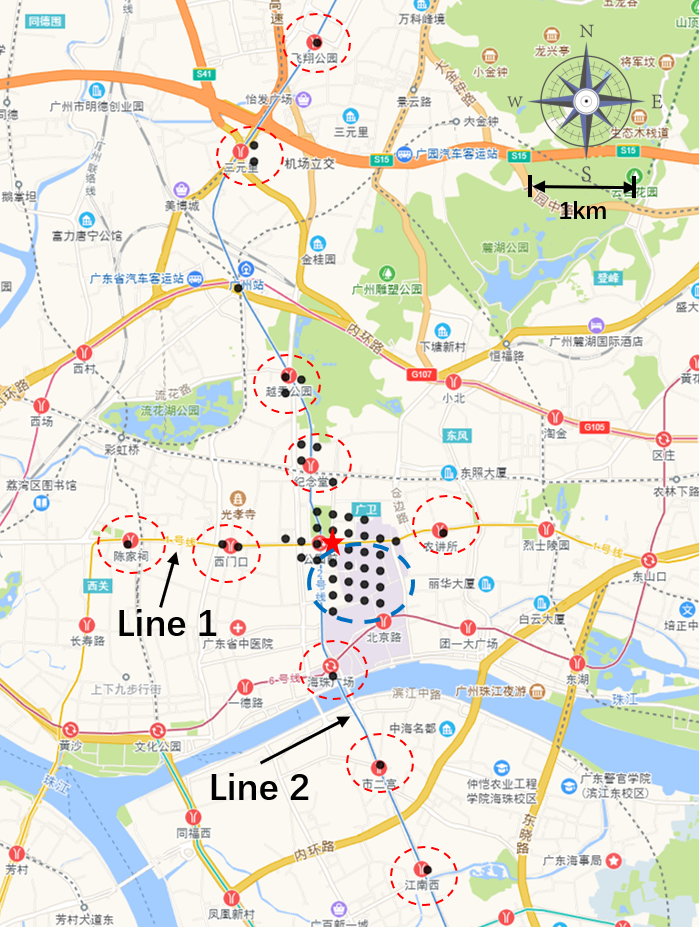}
         \caption{Gongyuanqian subway station (an interchange station of subway lines).}
         \label{gongyuanqian}
     \end{subfigure}
     \hfill
     \begin{subfigure}[t]{0.45\textwidth}
         \centering
         \includegraphics[width=\textwidth]{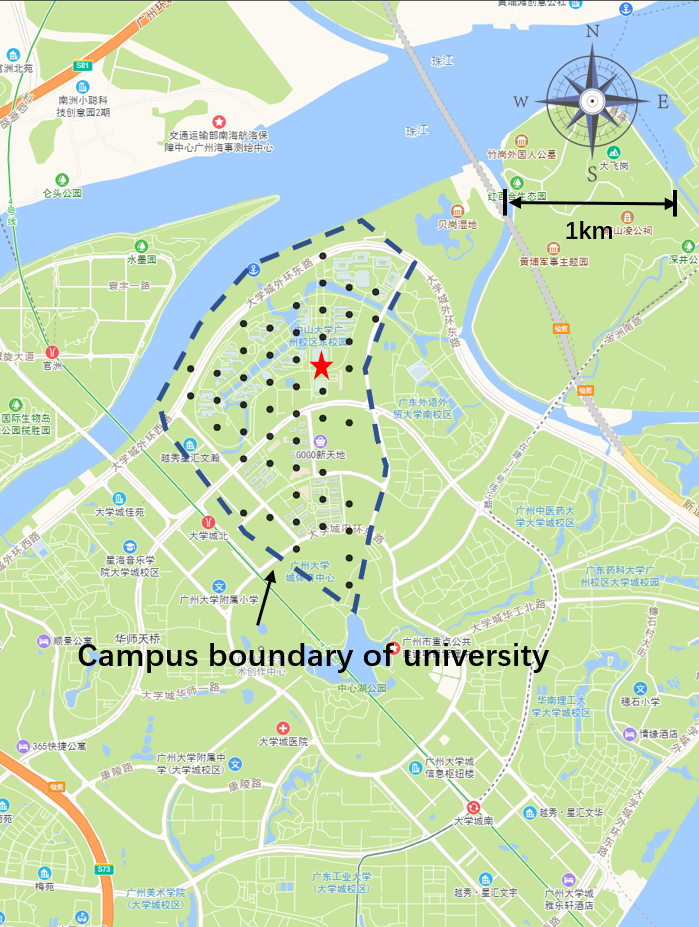}
         \caption{University campus in the High Education Mege Center.}
         \label{daxuecheng}
     \end{subfigure}
     
     \medskip

     \begin{subfigure}[t]{0.45\textwidth}
         \centering
         \includegraphics[width=\textwidth]{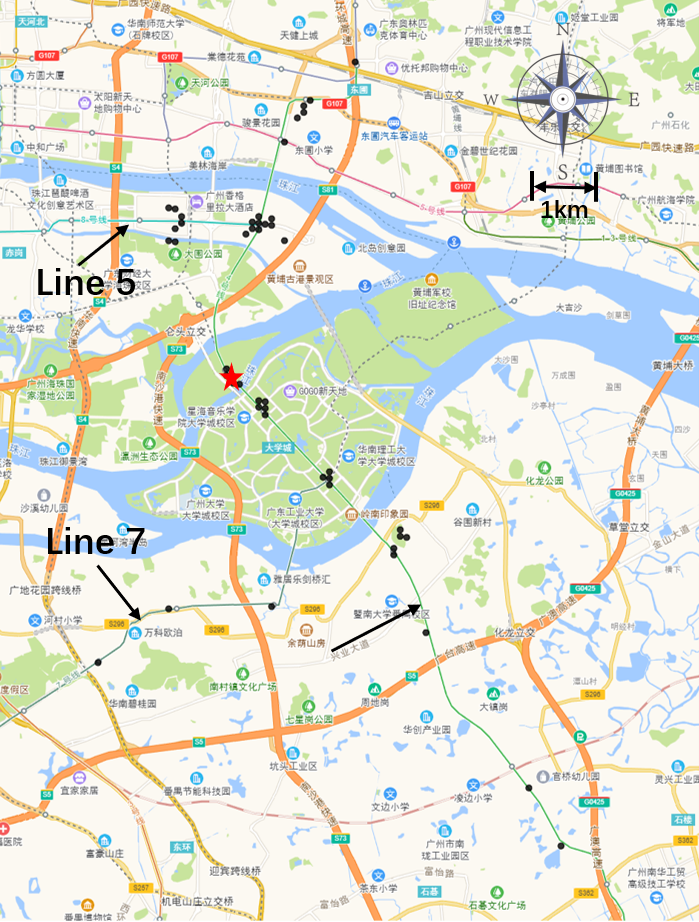}
         \caption{Guanzhou subway station (an important transportation hub).}
         \label{fig:guanzhou}
     \end{subfigure}
     \hfill
     \begin{subfigure}[t]{0.45\textwidth}
         \centering
         \includegraphics[width=\textwidth]{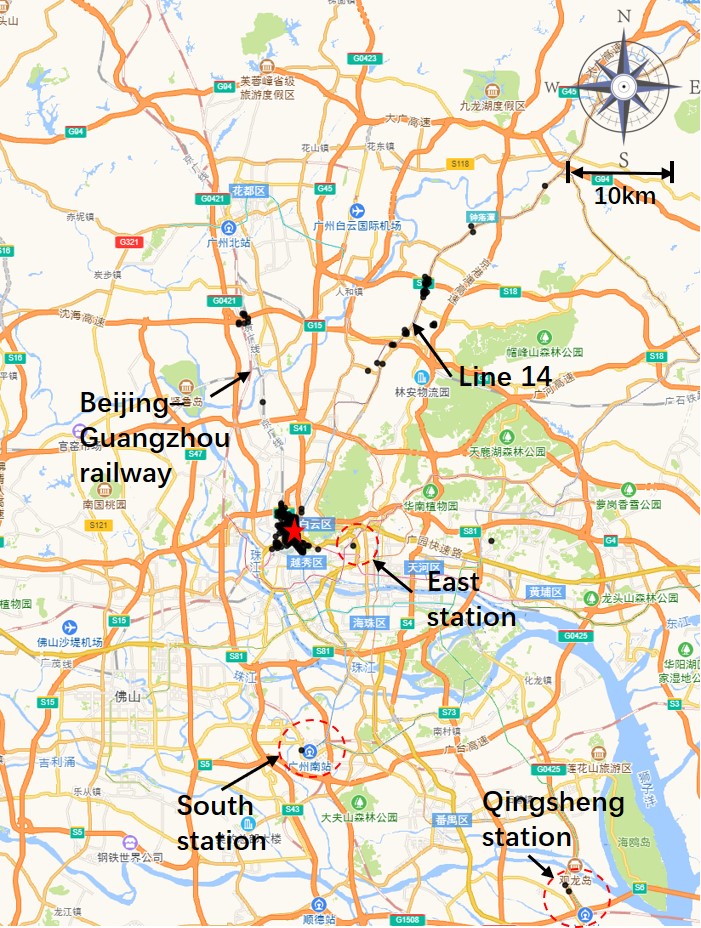}
         \caption{Guangzhou railway station.}
         \label{fig:trainstation}
     \end{subfigure}
     \hfill
        \caption{Representative locations and their related locations.}
        \label{fig:cases}
\end{figure}

There are also some popular locations. In Figure \ref{gongyuanqian}, select the Gongyuanqian subway station as the center location and calculate its top 50 related locations. The subway station is an interchange station of Line 1 (east-west direction) and Line 2 (north-south direction) of Guangzhou Metro. Overall, the distribution has a cross shape. It is observed that upstream and downstream subway stations along these two lines have a strong relationship with the center location, which are labeled in red dotted circles, indicating many people move among these stations by passing through Gongyuanqian station. Also, its nearby locations are also highly related. Especially the area labeled in the blue dotted circle, which is a popular commercial street called Beijing Road with many people moving around. 
In Figure \ref{daxuecheng}, we select the location of a college campus in the High Education Mega Center as the center and its top 50 related locations. It shows that most related locations are bounded within the campus. In other words, the embedding learns the boundaries simply using mobility data without any additional AOI information. It is reasonable because most students and teachers spend their time on the campus and less time outside.
In Figure \ref{fig:guanzhou}, we select the location of Guanzhou Station, which is one of the transportation hubs in Guangzhou. The top 50 most related locations are extended along transportation lines and most of the related locations are in remote places. Compared with the above two cases, there are fewer related locations nearby, because people usually do not spend much time in this location as a transfer station, and there are few POIs.
The Guangzhou Railway Station is located in the city center and is the most important railway station of this city for long-distance trips. In Figure \ref{fig:trainstation}, zoom in to a higher level and take a look at its top 200 most related locations. Except for the locations distributed in the city center nearby the station, the model can extract related locations that are centered in rather distant areas. For example, some related locations are in the Guangzhou South Railway Station, Guangzhou East Railway Station, and Qingsheng Station, because there are trains transporting between these stations.

All these cases show that the model encodes related locations to similar embedding vectors in a meaningful way. The results are reasonable and different types of locations show different distribution patterns.

\section{Conclusion}
This work tackles the location embedding problem and proposes a general-purpose space representation model in an unsupervised manner. For learning representations of discrete spatial regions, our method jointly embeds context information in human mobility and spatial adjacency and mitigates the data sparsity problem, especially in less populated areas. By this approach, we are able to capture relationships among locations and provide a better notion of semantic similarity in a spatial environment. Across quantitative experiments and case studies, we empirically demonstrate that the representations learned by GCN-L2V are effective and can be imported as features for down-streaming tasks. Our proposed method can be applied in a complementary manner to other place embedding methods and down-streaming GIS-related tasks. As far as we know, this is the first study that provides a fine-grained location embedding for every part on large scale using only LBS records. For future works, temporal information in LBS records should also be considered because human mobility is highly related to time periods. To boost related research in this area, we plan to implement a set of open-source location embedding of the whole country.


\bibliographystyle{splncs04}  
\bibliography{references}  






\end{document}